\documentclass{webofc}
\usepackage[varg]{txfonts}
\usepackage{graphicx} 
\usepackage{dcolumn}
\usepackage{bm}
\usepackage{amsmath,xcolor,subcaption,braket}

\begin{document}

\title{Adiabatic Hydrodynamization: a Natural Framework to Find and Describe Prehydrodynamic Attractors}
\author{\firstname{Krishna} \lastname{Rajagopal}\inst{1,2}\fnsep\thanks{\email{krishna@mit.edu}} \and
        \firstname{Bruno} \lastname{Scheihing-Hitschfeld}\inst{1}\fnsep\thanks{\email{bscheihi@mit.edu}} \and
        \firstname{Rachel} \lastname{Steinhorst}\inst{1}\fnsep\thanks{Speaker at QM23; \email{rstein99@mit.edu}}
}

\institute{Center for Theoretical Physics, Massachusetts Institute of Technology, Cambridge, MA  02139, USA 
\and
           CERN, Physics Department, TH Unit, CH-1211 Geneva 23, Switzerland
          }

\date{\today}

\abstract{
The adiabatic hydrodynamization framework~\cite{Brewer:2019oha,Brewer:2022vkq} is a promising framework within which to describe and characterize pre-hydrodynamic attractors in a model-independent fashion. Using this framework, we define a procedure to identify a time-dependent change in coordinates which reveals a dynamical reduction in the number of active degrees of freedom. Applying this procedure to the kinetic theory of a  Bjorken-expanding gas of gluons in the small angle elastic scattering limit, we are able to intuitively explain the self-similar evolution of the gluon distribution function long before the applicability of hydrodynamics, as well as the loss of memory of its initial condition.
}

\maketitle


Describing the far-from-equilibrium evolution of quark-gluon plasma (QGP) prior to the applicability of a hydrodynamic description is vital to understanding how the matter produced in relativistic nuclear collisions then rapidly acquires the features of a hydrodynamic fluid. Studies of this pre-hydrodynamic stage using various methods, including classical field simulations, kinetic theory, and holography, have discovered the existence of ``attractor" solutions to various simplified theories, as reviewed in  Ref.~\cite{Soloviev:2021lhs}. That is, the hydrodynamizing plasma is characterized by a rapid loss of sensitivity to the initial condition as it is driven towards a preferred region of phase space whose subsequent evolution can be characterized by a small number of quantities, far fewer than are needed to describe the myriad possible initial conditions that are all driven to the same attractor. In the context of QCD kinetic theory under Bjorken flow, provided the coupling is sufficiently weak, this has been understood through the presence of universal scaling in the evolution of gluon and quark distribution functions. For example, in the first stage of the bottom-up thermalization scenario~\cite{Baier:2000sb}, the gluon distribution function takes the self-similar form
\begin{equation} \label{eq:bmss}
    f(\bm{p},\tau) = \tau^\alpha w_S \left(p_\perp \tau^{-\beta},p_z \tau^{-\gamma}\right)
\end{equation}
where $\alpha = -2/3$, $\beta = 0$, and $\gamma = 1/3$ are called scaling exponents. It was recently found~\cite{Mazeliauskas:2018yef} that in QCD effective kinetic theory (EKT), a more general embodiment of this pre-thermal scaling form holds prior to relaxation onto Eqn.~\eqref{eq:bmss}, namely
\begin{equation} \label{eq:scaling}
	f(\bm{p},\tau) = \tau^\alpha(\tau) w_S \left(p_\perp \tau^{-\beta(\tau)},p_z \tau^{-\gamma(\tau)}\right)\,.
\end{equation}
In this ``pre-scaling regime'', the scaling exponents are time dependent, but knowledge of the few scaling exponents still suffices to describe the evolution of the full state of the system.

In Ref.~\cite{Brewer:2019oha}, the Adiabatic Hydrodynamization (AH) framework was introduced as a means to explain intuitively how and why   rapidly Bjorken-expanding QGP loses sensitivity to all but a small number of degrees of freedom. The AH hypothesis is that there exists a description of EKT dynamics that is analogous to Schr\"odinger time evolution with an effective Hamiltonian in which a small number of the lowest-``energy" modes are long-lived, while higher-``energy" modes quickly die away. This framework has been applied to the over-occupied limit of the small-angle scattering approximation and successfully explained the emergence of pre-thermal scaling ~\cite{Brewer:2022vkq}. In this work, we will relax the assumptions made in Ref.~\cite{Brewer:2022vkq}, in so doing demonstrating the robustness and generalizability of the AH framework.

\section{Adiabaticity and Scaling} \label{sec:adiabaticity}
We shall consider a homogeneous Bjorken-expanding gas of interacting gluons described in kinetic theory by the Boltzmann equation
\begin{equation}
	\frac{\partial}{\partial \tau} f(p_\perp,p_z,\tau) - \frac{p_z}{\tau} \frac{\partial}{\partial p_z}f(p_\perp,p_z,\tau)= -C[f(p_\perp,p_z,\tau)]\,.
\end{equation}
The gluon distribution function $f = f(p_\perp,p_z,\tau)$ can always be written in the form
\begin{equation} \label{eq:rescaledf}
	f(p_\perp,p_z,\tau) = A(\tau) w\left(\frac{p_\perp}{B(\tau)}, \frac{p_z}{C(\tau)},\tau \right) = A(\tau) w(\zeta,\xi,\tau)\,,
\end{equation}
where $A(\tau), B(\tau),$ and $C(\tau)$ are time-dependent momentum rescalings. In principle, $f$ can be written in this form for any choice of $A,B,C$, but in the scaling regime there exists a choice that eliminates the explicit time dependence in $w$, and the first derivatives of these scales,
\begin{equation}
	\alpha \equiv \frac{\tau}{A}\frac{\partial A}{\partial \tau}, \;\;\; \beta \equiv -\frac{\tau}{B}\frac{\partial B}{\partial \tau}, \;\;\; \gamma \equiv -\frac{\tau}{C}\frac{\partial C}{\partial \tau},
\end{equation}
can then be interpreted as scaling exponents.

Using this form, we can recast the Boltzmann equation into the form
\begin{equation} \label{eq:hamiltonianevolution}
	- \partial_y w = H_{\rm eff} w
\end{equation}
(where $y \equiv \log \tau/\tau_0$ is a convenient choice of time variable) because it turns the Boltzmann equation into a pseudo-Schr\"odinger equation in which $H_{\rm eff}$ is an operator that we will call the ``effective Hamiltonian". This operator will in general be a non-Hermitian and highly non-linear operator, but nonetheless we can understand the rescaled evolution equation~\eqref{eq:hamiltonianevolution} via the analogy with quantum mechanics. The presence of the sign -1 (rather than the usual $i$) in this pseudo-Schr{\"o}dinger equation has the important consequence that it allows us to identify the long-lived degrees of freedom of the system as the lowest energy mode(s) of the effective Hamiltonian. Specifically, if we express the rescaled state $w$ of the system as a linear combination of instantaneous eigenstates $\ket{n(y)}_R$ of $H_{\rm eff}$ with eigenvalues $\epsilon_n(y)$,
\begin{equation} 
    w(t) = \sum_n a_n(y) \ket{n(y)}_R = \sum_n \tilde{a}_n(y) e^{- \int^y dy' (\epsilon_n(y') } \ket{n(y)}_R \,, 
\end{equation}
one can show that the contribution to the superposition coming from all but the lowest-eigenvalue states decays as $\sim e^{-\epsilon_n y}$ if the evolution is adiabatic. That is, if
\begin{equation} \label{eq:coeffadiabaticity}
    \partial_y \log \frac{\tilde{a}_n}{\tilde{a}_m} \ll |\epsilon_n(y)-\epsilon_m(y)| \, .
\end{equation} 
The lowest-eigenvalue states can then be identified as long-lived modes. If the system is close enough to the ground state $\ket{0}_R$ of $H_{\rm eff}$, the 
condition~\eqref{eq:coeffadiabaticity} is approximately equivalent to~\cite{Brewer:2019oha} 
\begin{equation}
    \delta_A^{n} = \left| \frac{\bra{ n}_L\partial_y \ket{0}_R}{\epsilon_n-\epsilon_0} \right| \ll 1\,.
\label{AdiabaticityCondition}
\end{equation}
If $\partial_y \ket{0}_R = 0$, meaning that $\delta_A^{n}=0$, the adiabaticity condition \eqref{AdiabaticityCondition} is satisfied strictly; in this case, if $w$ has reached its ground state it is independent of $\tau$, and this condition is strictly equivalent to scaling.
The condition \eqref{AdiabaticityCondition} as written makes precise the notion of $\ket{0}_R$, the ground state of $H_{\rm eff}$, changing slowly and  adiabatically; when this condition is satisfied, approximate scaling is obtained.
%
%
In this way, the AH picture provides an intuitive understanding of how and why attractors arise and of their subsequent evolution in those kinetic theories to which it applies.
%
%
In the next Section, we take advantage of this picture by identifying an appropriate choice of rescalings $A(y),B(y),C(y)$ to minimize $\delta_A^n$, maximizing adiabaticity. In so doing, we shall demonstrate that the AH picture applies more broadly than was known previously and illuminate the underlying rescaled attractor behavior.

\section{Adiabaticity in the Pre-Thermal Attractor} \label{sec:expanding}

\begin{figure}[t]
\vspace{-0.1in}
\centering
\begin{subfigure}{0.48\textwidth}
    \includegraphics[width=\textwidth]{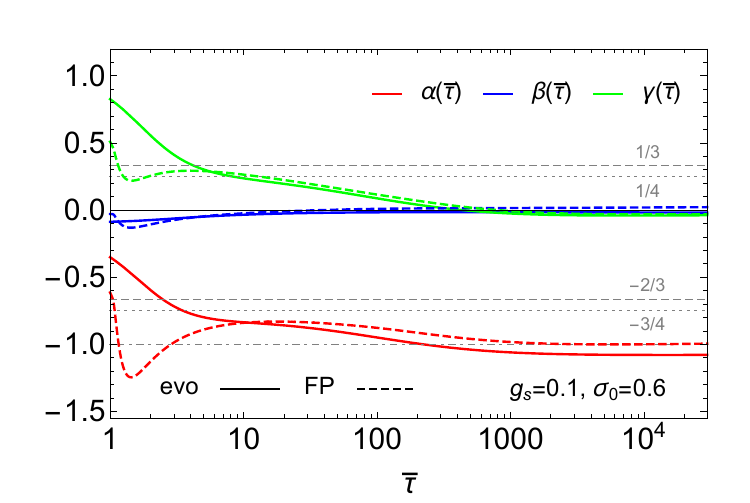} 
\end{subfigure}
\hfill
\begin{subfigure}{0.48\textwidth}
    \includegraphics[width=\textwidth]{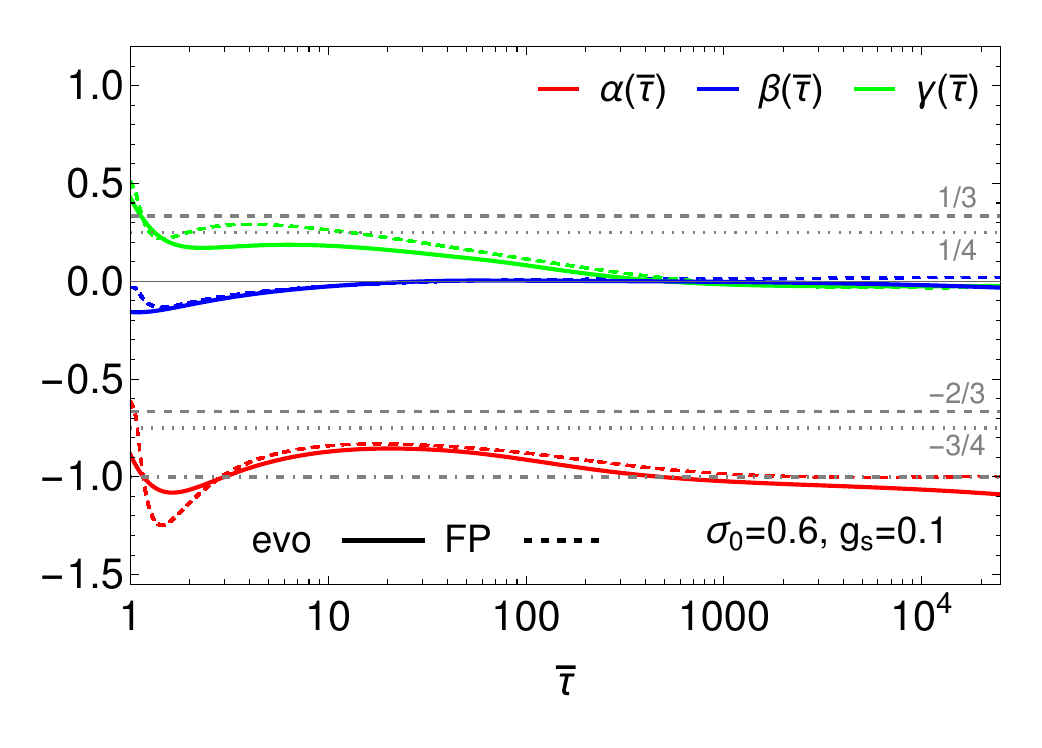} 
\end{subfigure}
\caption{
In these two panels, we display the results from three different calculations of the time evolution of the scaling exponents $\alpha$, $\beta$ and $\gamma$, all starting from the same initial conditions. The dashed curves in both panels show
scaling exponents calculated numerically in Ref.~\cite{Brewer:2022vkq} by averaging exponents extracted from moments of the Boltzmann equation, 
along the lines first done in Ref.~\cite{Mazeliauskas:2018yef}.
The left panel is taken from Ref.~\cite{Brewer:2022vkq}; the solid curves in this panel show scaling exponents found via a less generic variant of the adiabatic picture that applies only in a kinetic theory with a simpler version of the Fokker-Planck collision kernel than that we have used in this work. The solid curves in the right panel are the result of our calcualtion. They show the evolution of the scaling exponents found by optimizing adiabaticity at every time step via the procedure described in the text. 
}
\vspace{-0.1in}
\label{fig:scaling} 
\end{figure}

In our work~\cite{Rajagopal_2024}, we apply this framework to QCD EKT in the small-angle scattering limit, and consider only elastic scatterings; in this limit the collision kernel is ~\cite{Mueller_2000,Blaizot_2013}
\begin{equation}
	C[f] = - \left( \frac{g_s^4 N_c^2}{4\pi} \right) \log \left( \frac{p_{UV}}{p_{IR}} \right) \left[ I_a[f] \nabla^2_{\bm{p}} f + I_b [f]\nabla_{\bm{p}} \cdot \left( \frac{\bm{p}}{p} (1+f)f \right) \right],
\end{equation}
where 
\begin{equation}
	I_a \equiv \int \frac{d^3p}{(2\pi)^3} f (1+f), \;\;\; I_b \equiv \int \frac{d^3p}{(2\pi)^3} \frac{2}{p}f.
\end{equation}
We take the UV cutoff $p_{UV}$ to be $B(y)$ since this should represent the typical momentum scale, and take the IR cutoff $p_{IR}$ to be the Debye mass $m_D = \sqrt{2N_c g_s^2 I_b}$. We neglect for convenience the term proportional to $I_b f^2$ in $C[f]$. This kinetic theory is a generalization of Ref.~\cite{Brewer:2022vkq} in that we include all terms included in that work, plus the term proportional to $I_b \nabla \cdot \hat{p} f$. Upon casting $f$ into the rescaled form ~\eqref{eq:rescaledf}, we find the effective Hamiltonian
\begin{equation}
	H_{\rm eff} = \alpha + \beta \zeta \partial_\zeta + (\gamma-1) \xi \partial_\xi - q \left[ \frac{1}{B^2} \left( \frac{1}{\zeta} \partial_\zeta + 			\partial_\zeta^2 \right) + \frac{1}{C^2} \partial_\xi^2 \right] - \frac{\lambda}{p} \left( 2 +  \zeta \partial_\zeta +  \xi \partial_\xi \right).
\end{equation}
Let us also make for convenience the approximation $p \approx p_\perp = B\zeta$. Assuming an appropriate choice of time-dependent rescalings, the system's evolution can be approximately expressed in terms of a convenient set of left and right basis states,
\begin{equation}
	\psi_{ij}^{(R)} = p_i(\zeta) q_j(\xi) e^{-\left(\xi^2/2+\zeta\right)}, \;\; \psi_{ij}^{(L)} = p_i(\zeta) q_j(\xi),
\end{equation}
allowing us to write the effective Hamiltonian in matrix form,
\begin{equation}
	H^{ik}_{jl} = \braket{\psi^L_{ij} | H | \psi^R_{kl}} = \int \frac{d^2\zeta d\xi}{(2\pi)^3} \psi^L_{ij} H \psi^R_{kl}
\end{equation}
Given this matrix form of $H_{\rm eff}$, we can at each time solve for its instantaneous eigenstates and eigenvalues. Let us then choose $B(y),C(y)$  at each point so as to minimize $\braket{\partial_y \ket{0}_R | \partial_y \ket{0}_R}$, and choose $A(y)$ to set the ground state energy to zero. 
This differs from the work done in Ref.~\cite{Brewer:2022vkq} in that here no analytic expressions for the eigenstates are known, and adiabaticity is not exactly satisfied. However, both here and in Ref.~\cite{Brewer:2022vkq}, the adiabatic framework is able to reproduce scaling exponents found by calculating moments of the Boltzmann equation; this is shown in 
Fig.~\ref{fig:scaling}. Our results, illustrated by the solid curves in the right panel of Fig.~\ref{fig:scaling}, demonstrate that the AH picture provides a quantitative description of the pre-hydrodynamic attractor
in a more general kinetic theory, 
which has required us to 
optimize adiabaticity in a context in which it is not guaranteed.

AH is a promising framework for intuitively explaining how and why scaling and attractor behavior arises long before hydrodynamization. In this work we have demonstrated the robustness of the connection between pre-thermal scaling behavior and adiabatic hydrodynamization, generalizing its applicability. The AH setup used in these proceedings has the remaining limitation of assuming $p \approx p_\perp$, which prevents the use of this setup once the system begins to isotropize. In upcoming work~\cite{Rajagopal_2024}, we will expand the application of this framework to a yet more general scenario, allowing us to reproduce the scaling seen in every phase of the bottom-up scenario~\cite{Baier:2000sb}, connecting the early-time scaling behavior to hydrodynamization seamlessly, with the entire evolution described adiabatically.

\bibliography{main}

\end{document}